\documentclass{amsart}

\usepackage{amssymb,amsmath}
\usepackage{graphics}

\newcommand{\tr}{\mathop{\mathrm{tr}}\nolimits}
\newcommand{\ii}{\mathrm{i}}
\newcommand{\pd}{\partial}

\newcommand{\A}{\mathcal{A}}
\newcommand{\B}{\mathcal{B}}

\newcommand{\F}{{\mathcal F}}

\newcommand{\hh}{\mathcal{H}}
\newcommand{\X}{X^{(0)}}
\title[Equivalence\&Condensation]{On the Equivalence of Noncommutative Models in Various
Dimensions and Brane Condensation}
\author{Corneliu Sochichiu}
\address{The Abdus Salam ICTP, strada Costiera 11, 34100 Trieste, Italy}
\address{Bogoliubov Laboratory of
Theoretical Physics, Joint Institute for Nuclear Research, 141980
Dubna, Moscow Reg., RUSSIA}
\address{Institutul de Fizic\u a Aplicat\u a
A\c S, str. Academiei, nr. 5, Chi\c sin\u au, MD2028 MOLDOVA}

\begin{document}
\begin{abstract}
Here we construct a map from the algebra of fields in
two-dimen\-sio\-nal noncommutative of U(1) Yang--Mills fields
interacting with Kaluza--Klein scalars to a $D$-dimensional one,
as a solution in the two-dimensional model. This proves the
equivalence of noncommutative models in various (even)
dimensions. Physically this map describes condensation of
D1-branes.
\end{abstract}
\maketitle
\section{Introduction}

Noncommutative models appear to be relevant to the description of
various aspects of string theory
\cite{Witten:2000nz,Seiberg:1999vs}.

An approach to noncommutative gauge theories can be developed using
matrix models \cite{Ishibashi:1996xs,Banks:1997vh} in the limit of
large matrices.

Compactifications of these models to noncommutative tori was shown to
yield the noncommutative Yang--Mills models \cite{Connes:1998cr}. The
noncommutative tori correspond to matrix configurations in the IKKT
model having \emph{nondegenerate} scalar commutator in the limit
$N\to\infty$. Although, at infinite matrix size these configurations
solve equations of motion, they do not correspond to local minima of
the classical action, and therefore do not contribute at the level of
perturbation theory. However, due to their large entropical factor
these configurations should become important in the strong coupling
regime.

In the case of finite $N$ one can construct a map from the matrix
model to some kind of non-commutative lattice gauge model (for a
recent review see \cite{Ambjorn:2000cs} and references therein).

The limit $N\to\infty$ was studied by the author in Refs.
\cite{Sochichiu:2000ud,Sochichiu:2000fs}, where it was shown that
this limit is ambiguous at least in the perturbative approach.
The ambiguity consists in the fact that depending on the
background solution chosen one has in the $N\to\infty$ limit
either ten-dimensional  Yang--Mills--type model or its reductions
to lower dimensions.

In what follows we will start with a particular choice from the
variety of possible models arising in the $N\to\infty$ limit of
the IKKT matrix model. We are going to show that in fact there is
a wide universality among these models, in particular, matrix
fluctuation around a $D$-dimensional (commutative) solution are
completely equivalent to perturbations around the mentioned
configuration having nondegenerate scalar commutator in $2D$
dimensions and corresponding to noncommutative tori. This
equivalence become manifest due to the possibility to absorb the
kinetic term in noncommutative gauge models.\footnote{I learned
this possibility from Ref. \cite{Gopakumar:2000zd}.}

Also, we explicitly build a one-to-one map from one-dimensional
noncommutative gauge model to the $D$-dimensional one using the
isomorphism of Hilbert spaces. The $D$-dimensional model can be
implemented as a solution in the one-dimensional case while the
last can be obtained through the dimensional reduction of the
former.

The plan of the paper is as follows. First we review some results
concerning IKKT model, after that we describe the equivalence
between fluctuations around commutative/noncommutative solutions,
and finally find the solution which gives the map from two- to
the $D$-dimensional noncommutative models, and discuss the
implications of this map.

For the notations and background of this work we refer the reader
to the Ref.~\cite{Sochichiu:2000ud}.

\section{$N=\infty$ IKKT model}
Start with the IKKT model at finite $N$. It is given by the
classical Euclidean action,
\begin{equation}\label{act_ikkt}
    S_{\mathrm{IKKT}}=-\frac{1}{4g^2}\tr[X_\mu, X_\nu]^2-\bar{\psi}
    \Gamma^\mu[X_\mu,\psi],
\end{equation}
where $X_\mu$, and $\psi$ are scalar and spinor matrices with
large size $N\to\infty$. Note that in this paper the Greek labels
always run in ten dimensions, $\mu=1,\dots,10$. This model
possesses a number of properties such as supersymmetry,
SO(10)-Lorentz and SU(N)-gauge invariances
\cite{Ishibashi:1996xs}.

At finite $N$, the only classical solutions to this model are
given by sets of commuting matrices $\X_\mu$,
\cite{Sochichiu:2000ud},
\begin{equation}\label{pp=0}
    [\X_\mu,\X_\nu]=0.
\end{equation}

If eigenvalues of matrices $\X_\mu$ form a $D$-dimensional
lattice, where $D$ is an arbitrary integer in the range, $0\leq
D\leq 10$, then they can be expressed as functions of $D$
independent matrices $p_i$, where the Latin indices run through
$D$ dimensions, $i=1,\dots,D$.

In the limit $N\to\infty$ matrix fluctuations around such a
background are described by a $D$-dimensional Yang--Mills type
model. In particular, when $D=10$, and matrices $\X_\mu$ are
independent one can identify them with $p_i$. (In this case the
sets of Greek and Latin indices coincide.) The limiting $N=\infty$
model in this case is equivalent to one given by the following
action, \cite{Sochichiu:2000ud},
\begin{equation}\label{s_func}
  S=-\int d^{D}x\, d^{D}l
  \left(\frac{1}{4}\F_{\mu\nu}^2(x,l)+
  \ii\bar{\psi}\Gamma^\mu\nabla_\mu
  \psi(x,l)\right),
\end{equation}
where,
\begin{align}\label{F}
  &\F_{\mu\nu}={\pd_\mu}\A_\nu(x,l)-
  {\pd_\nu}\A_\mu(x,l)
  +g[\A_\mu,\A_\nu]_{\star}(x,l) \\ \label{nabla}
  &\nabla_\mu\psi(x,l)={\pd_\mu}\psi(x,l)
  -g[\A_\mu,\psi]_{\star}(x,l).
\end{align}
Here derivatives are computed with respect to $x$, and they are
given by,
\begin{equation}\label{deriv}
   \pd_\mu \A(x,l)\equiv \ii [\X_\mu (l),
   \A(x,l)],
\end{equation}

Commutators in eqs. (\ref{s_func}--\ref{deriv}) are computed using
the star product,
\begin{equation}\label{scom}
  [\A,\B]_{\star}=\A\star \B-\B\star \A,
\end{equation}
while the star product itself is defined as,
\begin{equation}\label{prod->*}
  \A\star \B(x,l)=\left.e^{-\frac{\ii}{2}
  \left(\frac{\pd^2}{\pd x' \pd l}-\frac{\pd^2}{\pd x\pd l'}\right)}
  \A(x,l)\B(x',l')\right|_{x'=x \atop l'=l}.
\end{equation}
Here $l_i$ (the same as $l_\mu$) parameterise the spectrum of
$p_i$, while $x^i$ are coordinates on the (Fourier) dual space
\cite{Sochichiu:2000ud}.

In the case when $D<10$ the model is given by the reduction of
(\ref{s_func}) to $D$ dimensions. In this case, $x^i$ and $l_i$
are $D$-dimensional and the set of Greek indices comes split in
two subsets formed by Latin indices $i,j,k\dots$ and other one of
Kaluza--Klein multiplet indices which appear upon reduction to
$D$ dimensions from 10 dimensions. We will not introduce the last
type of indices, because along this paper we do not need the this
split explicitly, but keep instead the Greek letters for both
space-time and Kaluza--Klein multiplets. In this case the
derivative terms generalise according to (\ref{deriv}), and the
dimensional reduced action keeps the same form as given by eqs.
(\ref{s_func}--\ref{prod->*}), but one should keep in mind that
$\A_\mu(x,l)$ in this case denote both the $D$-dimensional gauge
field and the Kaluza--Klein scalars.

Using the Moyal correspondence, the algebra of $(x,l)$-functions
supplied with the star product (\ref{scom}) can be seen as a
representation of the $D$-dimensional Heiseberg algebra with the
Weyl ordering prescription which is generated by $l_i$ and $x^i$,
acting on the Hilbert space $\hh_{D}$ and satisfying the
commutation relation,
\begin{equation}\label{canon}
  [l_i,x^j]=-\ii \delta_i^j.
\end{equation}
In this case the integration over $d^Dx\, d^Dl$ is equivalent to
taking the trace over $\hh_{D}$. In what follows we will not
distinguish between these two forms.

By a redefinition of field $\A_\mu$,
\begin{equation}\label{redef}
   \A_\mu\to \X_\mu(l_i)+\A_\mu,
\end{equation}
one can absorb the kinetic term in eq. (\ref{s_func}). As a result one
has the model described by the action (in Heisenberg form),
\begin{equation}\label{s_op}
   S=-\tr_{\hh_D}\left( \frac{1}{4}[\A_\mu,\A_\nu]^2+
   \bar{\psi}\Gamma^\mu[\A_\mu,\psi]\right),
\end{equation}
where $\A$ and $\psi$ are hermitian operators of $D$-dimensional
Heisenberg algebra which act on $\hh_D$ and are represented by
noncommutative functions on $(l_i,x^i)$, the trace
$\tr_{\hh_D}(\cdot)$ denotes the integration over noncommutative
phase space, $\int d^Dxd^Dl(\cdot)$. Once again note that
$\mu,\nu,\dots=1,\dots, 10$, while the fields are defined on
$2D$-dimensional noncommutative space generated by $l$ and $x$.

Let us note that in the form (\ref{s_op}) the model is manifestly
invariant with respect to reparameterisations of $(l,x)$
preserving the commutator (\ref{canon}).

Although, the action (\ref{s_op}) or its dimensional reductions
are obtained as a continuum limit of fluctuations around a
commutative background ($[\X_\mu,\X_\nu]=0$), one can show that
the same model can be obtained as a continuum limit of
fluctuations around a configuration with $[\X_\mu,\X_\nu]=\ii
B_{\mu\nu}\neq 0$, in $2D$ dimensions. In other words, the model
(\ref{s_op}) is equivalent to the U(1) non-commutative
Yang--Millls model in $2D$ dimensions \cite{Connes:1998cr},  when
$2D=10$, or its $2D$ dimensional reduction when $2D<10$.

Indeed, consider action (\ref{s_op}) in the case of $D=5$. After
shifting back the fields $\A_\mu \mapsto (\A_\mu-\tilde{l}_\mu)$,
where $\tilde{l}_\mu$ are given by $\tilde{l}_i\equiv l_i$ for
$\mu=i=1,\dots,5$ and $\tilde{l}_{5+i}=x_i$, for
$\mu=5+i=6,\dots,10$ the action (\ref{s_op}) becomes one of the
10-dimensional noncommutative U(1) Yang--Mills
model.\footnote{Or, oppositely, absorbing the kinetic term, one
can bring the ten dimensional noncommutative U(1) Yang--Mills
model to the form (\ref{s_op}). This correspondence is possible
since $D$-dimensional Heisenberg algebra coincides with
$2D$-dimensional noncommutative space.} The same trick can be
made for any $D\leq 10$, however, the meaning of the model when
$20\geq 2D>10$ is not yet clear, but as we are going to
demonstrate later this is not a problem since any $D>1$ model is
equivalent to the $D=1$.

\section{From one dimension up to $D>1$}
In this section we consider the model given by the action
(\ref{s_op}) with $D=1$, which means that the fields are defined
as functions on the one-dimensional Heisenberg algebra generated
by $l$ and $x$, satisfying usual commutation relation,
\begin{equation}
   [l,x]=-\ii\hbar,
\end{equation}
which is the same as two-dimensional noncommutative space. Let us
note that the one-dimensional solution is one having the largest
entropical factor \cite{Sochichiu:2000ud} in the IKKT matrix
model.

For convenience consider the Heisenberg algebra to be defined on a
circle: $x+2\pi L\sim x$. In this case the momentum operator $l$
have discrete spectrum. Its eigenvalues are given by $n/L$,
$n=0,\pm 1,\pm 2, \dots$. Later one can take the limit $L\to
\infty$.

Consider the equations of motion for this model. Vacuum solutions
(with $S=0$), are given by the commutative sets of operators
$\A^{(0)}_\mu$. In terms of ``smooth'' functions and up to a
gauge transformation, they are given by the arbitrary functions
$\A^{(0)}_\mu(l)$. As it is generally known, two and  more
\emph{continuous} functions of one variable always form a
functional dependent set, therefore, the set of $\A^{(0)}_\mu(l)$
is a such one if $\A^{(0)}_\mu(l)$ are continuous functions of
$l$. This property, however does not hold for the discontinuous
functions.

After the kinetic term absorption the equations of motion are no
more differential equations, thus the condition imposed one the
solutions of equations of motion to be smooth functions are no
more justified and can be given up. Moreover, the spectrum of $l$
itself is discrete.

From the other hand, since the smoothness properties are strictly
related to the notion of topology of the space-time, the
non-smooth solution can be interpreted as the changing of the
space-time topology. The space-time topology can be extracted
from the set of operators $\A^{(0)}_\mu$ in the framework of
Connes' approach \cite{Connes:2000by}.

Consider that the solution $\A^{(0)}_\mu$ now carries the
attributes of the $D$ dimensional space, i.e. they form a
$D$-dimensional lattice, and can be expressed as functions of the
basic set of independent operators $l_i$, $i=1,\dots,D$.

In what follows let us construct an explicit solution with
functionally independent (and, therefore, discontinuous) $l_i(l)$.

Since eigenvalues of $l$ and $l_i$ form, respectively,
one-dimensional and $D$ dimensional lattices, the solution
$l_i(l)$ is given explicitly by the map from the one-dimensional
lattice $\Gamma_1$ of the eigenvalues of $l$ to the
$D$-dimensional lattice $\Gamma_D$ of eigenvalues of $l_i$. Due to
the reparameterisation invariance one can take both lattices to
be regular and rectangular ones.

The map can be constructed through the following sequence of steps
(the particular case for $D=2$ is depicted in the figure).
\begin{figure}
\scalebox{0.2}{\includegraphics{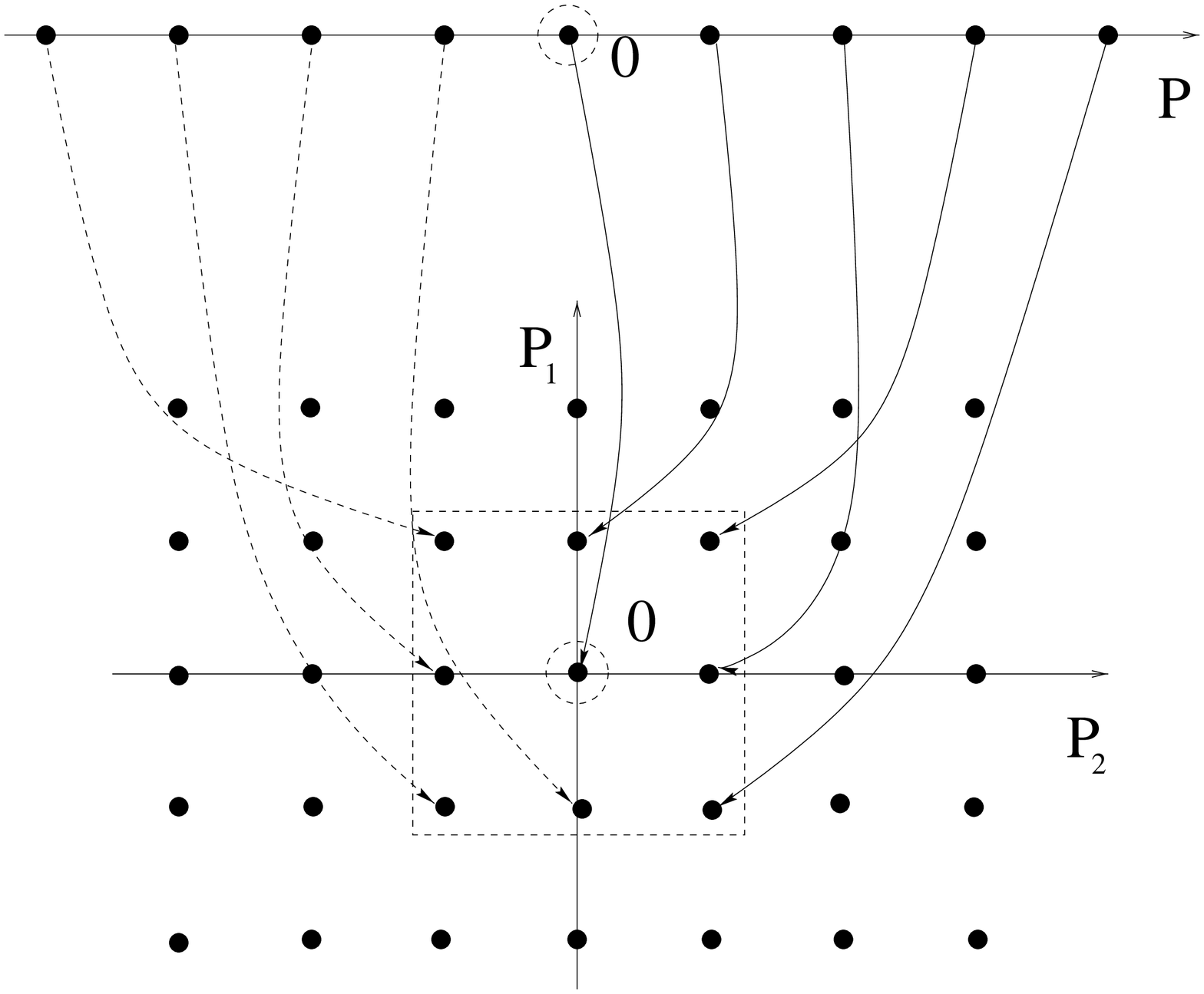}}
\caption{The isomorphic map from 1-dimensional lattice to the
2-dimensional one. It is shown how the first hyper-cubic shell (in the
dotted box) of $\Gamma_2$ is filled by the points of $\Gamma_1$.}
\end{figure}

\begin{enumerate}
  \item Map the origin of $\Gamma_1$ to the origin of $\Gamma_D$.
  \item Map $D$ points next to origin in the positive direction and
  $D$ ones in negative direction of $\Gamma_1$ to the nearest neighbor
  points to the origin of $\Gamma_D$, using e.g. lexicographic
  ordering.
  \item Fill the the remaining points of the $D$-dimensional hypercube
  of two lattice unites size centered at origin of $\Gamma_D$ by
  images of $n=\pm
  (D+1),\pm (D+2),\dots$ points of $\Gamma_1$.
  \item In the same manner fill the next hyper-cubic shell of
  $\Gamma_D$ by images of of points of $\Gamma_1$, etc.
  \item Thus, $(2n+1)^D$ points around the origin of $\Gamma_1$ fill the
  hyper-cube of the size $2n$  centered at the origin of $\Gamma_D$.
\end{enumerate}

As it can be seen by the construction the map is one-to-one.
However, the resulting operators $l_i$ are functionally
independent. Under this map small momenta are transfered to small
momenta, and large ones to large ones. This means that
respectively low/high energy states of one model are mapped into
low/high energy sector of another one.

This correspondence allows one to pass from the one dimensional
Heisenberg algebra to the Heisenberg algebra in arbitrary
dimension $D>1$ (or from two-dimensional noncommutative space to
$2D$ dimensional one).

Indeed, using the approach of Ref. \cite{Sochichiu:2000fs} one can
introduce for each operator $l_i$ its canonical conjugate $x^i$,
satisfying the Heisenberg commutation relation in $D$ dimensions,
\begin{equation}\label{d_heis}
  [l_i,x^j]=-\ii\delta_i^j.
\end{equation}

One can, therefore, pass from (\ref{s_op}) with $D=1$ to the
equivalent description as a model in a different dimension $D>1$.

Let us shortly describe the construction
\cite{Sochichiu:2000ud,Alvarez-Gaume:2000dx}. Consider the
eigenvalue problem for adjoint operators $P_i=[l_i,\cdot]$ and
$Q^i=[x^i,\cdot]$. This problem is consistent since $P_i$ and
$Q^j$ are commutative and self-adjoint on the space of square
integrable operators with bounded $P_i$. The eigenvalue problem
is solved by the (eigen)operators,
\begin{equation}
   E(k,z)=e^{\ii k_i q^i+\ii l_i z^i},
\end{equation}
where $k_i$ and $z^i$ are eigenvalues of $P_i$ and $Q^i$
respectively. Since, by the construction, both $(l,x)$ and
$(l_i,x^i)$ can be represented on the same Hilbert space $\hh_1$,
(in fact, we introduced isomorphism between $\hh_1$ and $\hh_D$),
one can expand an arbitrary square integrable operator $\A(l,x)$
of original model in the $E(k,z)$ basis, using the trace over the
Hilbert space of the one-dimensional Heisenberg algebra, and get
an operator $\A(l_i,x^i)$ in the $D$-dimensional model,
\begin{equation}\label{D_expand}
   \A(l_i,x^i)=\sum_{k_i,z^i} \tilde{\A}(k_i,z^i) E(k_i,z^i),
\end{equation}
where,
\begin{equation}
   \tilde{\A}(k_i,z^i)=\frac{1}{(2\pi)^D}\tr_{\hh_1}E^\dag (k_i,z^i)\A(l,x).
\end{equation}

Applying this procedure to the fields in model (\ref{s_op}) with
$D=1$, one obtains the equivalent description in terms of a $D$
dimensional model with any  $D>1$ .

Formally, we have introduced here a noncommutative (and
discontinuous) change of variables. Indeed, since $l_i$ and $x^i$
are invertible functions of one-dimensional $l$ and $x$, one can
find their inverse, $l=l(l_i,x^i)$ and $x=x(l_i,x^i)$, and plough
this dependence in the one-dimensional operator $\A(l,x)$ to get
the function $\A(l_i,x^i)=\A(l(l_i,x^i),x(l_i,x^i))$ which is a
$D$ dimensional operator.

Now we can give the following physical implication of this
construction.

The $D$-dimensional noncommutative gauge model describes, in
fact, the a D$p$-brane where $p=D-1$ \cite{Seiberg:1999vs}. We
constructed a solution in the two-dimensional noncommutative
gauge model which has the meaning of $2D$ dimensional space. This
solution gives the correspondence between the gauge models (gauge
fields interacting with Kaluza--Klein scalars) in various even
dimensions. Taking into account the brane interpretation of the
noncommutative gauge models with scalar fields, this describes the
condensation of D1-branes to a D$p$-brane, where $p=2D-1$. From
this point of view the multiple vacua of the IKKT model Ref.
\cite{Sochichiu:2000ud}, are nothing else than the condensation
of D(-1)-branes described by the IKKT matrices to an arbitrary
IIB brane (i.e. a brane with even-dimensional world-sheet), which
is the $2D$-dimensional reduction of the ten-dimensional
noncommutative U(1) gauge model.

\section{Conclusions}
Let us briefly summarise the results of this note.

First, we have shown that the model desribing the continuum limit
of fluctuations of the IKKT matrix  model around a commutative
background, $[p_i,p_j]=0$, in $D$ dimensions is equivalent to one
describing fluctuations around a $2D$ dimensional background
satisfying, $[p_I,p_J]=\ii B_{IJ}$, where $B_{IJ}$ is a
nondegenerate antisymmetric scalar matrix. ``Physically'' this
means that in the singular limit $B\to 0$ corresponds to doubling
of the dimensionality of the noncommutative gauge model.

Second, we demonstrated that the noncommutative model in $2$
dimensions can be isomorphically mapped to the $2D$-dimensional
one. The last feature may be interpreted as the noncommutative
geometry counterpart of the duality relating various branes or
their condensation,
\cite{Banks:1997vh,Ganor:1997zk,Banks:1997nn,Dijkgraaf:1997ku}.

In early works on IKKT model \cite{Ishibashi:1996xs}, it was
conjectured that this in the limit $N\to\infty$ generates the
space-time as the set of expectation values of operators $X_\mu$.
In this context it seems natural that the topology and, in
particular, the space-time dimensionality are also generated by
the solution $l_i(l)$.

\subsection*{Acknowledgements}
I benefited from useful discussions with many people. In particular, I
would acknowledge ones with E. Kiritsis and Yu. Makeenko.

This work was made under auspices of FSA program: ICTP--BLTP with
additional support from the High Energy Section of ICTP, it is a
pleasure to thank its director Prof. S. Randjbar-Daemi.

This work was supported by RFBR grant no.99-01-00190, INTAS grant
no.950681, Scientific School Support grant no.96-15-0628.

\end{document}